\documentclass[12pt]{article}

\begin{document}

\setcounter{page}{0} \thispagestyle{empty}

\vskip 2cm

\begin{center}
{\Large \textbf{Three-loop universal anomalous dimension of the\\[5mm]
Wilson operators in ${\mathcal N}=4$ SUSY Yang-Mills model }} \\[5mm]
\vspace*{10mm} A.~V.~Kotikov$^{a}$, L.~N.~Lipatov$^{b}$,
A.~I.~Onishchenko$^{c,d}$ and V.~N.~Velizhanin$^{b}$\\[10mm]

${}^{a}$ Bogoliubov Laboratory of Theoretical Physics \\[0pt]
Joint Institute for Nuclear Research\\[0pt]
141980 Dubna, Russia \\[0pt]
\vspace*{0.5cm} ${}^{b}$ Theoretical Physics Department, \\[0pt]
Petersburg Nuclear Physics Institute \\[0pt]
Orlova Rosha, Gatchina, \\[0pt]
188300, St. Petersburg, Russia\\[0pt]
\vspace*{0.5cm} ${}^{c}$ Department of Physics and Astronomy \\[0pt]
Wayne State University, Detroit, MI 48201, USA \\[0pt]
\vspace*{0.5cm} ${}^{d}$ Institute for Theoretical and Experimental Physics,
\\[0pt]
Moscow, Russia
\end{center}

\vspace{1cm}

\begin{center}
\textbf{Abstract}
\end{center}

We present results for the three-loop universal anomalous dimension $\gamma _{uni}(j)\,$
of Wilson twist-2 operators in the ${\mathcal N}=4$ Supersymmetric Yang-Mills model.
These expressions are obtained by extracting the most complicated contributions
from the three loop non-singlet anomalous dimensions in QCD which were
calculated recently. Their singularities at $j=1$ coincide with the predictions obtained from the
BFKL equation for ${\mathcal N}=4$ SYM in the next-to-leading order. The asymptotics of
$\gamma _{uni}(j)$ at large $j$ is in an agreement with the expectations
based on an interpolation between weak and strong coupling regimes in the
framework of the AdS/CFT correspondence.

\newpage

{\bf Introduction}
\vskip 3mm
The anomalous dimensions (AD) of the twist-two Wilson operators govern the
Bjorken scaling violation for parton distributions in a framework of Quantum
Chromodynamics (QCD)~\cite{LONLOAD}. These quantities are
expressed through the Mellin transformation
\[
\gamma _{ab}(j)=\int_{0}^{1}dx\,\,x^{j-1}W_{b\rightarrow a}(x)
\]
of the splitting kernels $W_{b\rightarrow a}(x)$ for the
Dokshitzer-Gribov-Lipatov-Altarelli-Parisi (DGLAP) equation~\cite{DGLAP}
which relates the parton densities $f_{a}(x,Q^{2})$ (hereafter $a=\lambda,\,g,\,\phi$ for
the spinor, vector and scalar particles, respectively) with different
values of $Q^{2}$ as follows
\[
\frac{d}{d\ln {Q^{2}}}f_{a}(x,Q^{2})=\int_{x}^{1}\frac{dy}{y}
\sum_{b}W_{b\rightarrow a}(x/y)\,f_{b}(y,Q^{2})\,.
\]
The
anomalous dimensions and splitting kernels in QCD are well known up to the
next-to-leading order (NLO) of the perturbation theory~\cite{LONLOAD}.

The QCD expressions for AD can be transformed to the case of the
${\mathcal N}=1$ Supersymmetric Yang-Mills theories (SYM)
if one will use for the Casimir operators $C_{A},C_{F},T_{f}$ the following
values $C_{A}=C_{F}=N_{c}$, $T_{f}=N_{c}/2$ (the last substitution follows
from the fact, that each gluino $\lambda_{i}$ being a Majorana particle gives a
half of the contribution for the Dirac spinor). For extended supersymmetric
theories the
anomalous dimensions cannot be obtained in this simple way, because
additional contributions coming from scalar particles should be also taken
into account~\cite{KL01}. Recently these
anomalous dimensions were calculated in the next-to-leading approximation~\cite{KoLiVe}
for the ${\mathcal N}=4$ Supersymmetric Yang-Mills theory.

It turns out, that the expressions
for eigenvalues of the AD matrix in the ${\mathcal N}=4$ SYM can be derived directly
from the QCD anomalous dimensions without tedious calculations by using a
number of plausible arguments. The method elaborated in Ref.~\cite{KL} for
this purpose is based on special properties of solutions of the
Balitsky-Fadin-Kuraev-Lipatov (BFKL) equation~\cite{BFKL,next} in this model
and a new relation between the BFKL and DGLAP equations (see~\cite{KL01}).
In the NLO approximation this method gives the correct results for
AD eigenvalues, which was checked by direct calculations in~\cite{KoLiVe}.
Its properties will be reviewed below only shortly and
a more extended discussion can be found in~\cite{KL}.

Next-to-next-to-leading order (NNLO) corrections to AD in QCD
were calculated recently~\cite{VeMoVo}\footnote{see also Ref.~\cite{VeMoVo2} for the singlet case}
in the nonsinglet case. Using these
results and the method of Ref.~\cite{KL} we derive in this paper the
eigenvalues of the anomalous dimension matrix for the ${\mathcal N}=4$ SYM in the NNLO
approximation.
\vskip 3mm
{\bf Evolution equation in ${\mathcal N}=4$ SYM}
\vskip 3mm
The reason to investigate the BFKL and DGLAP equations in the case of
supersymmetric theories is based on a common belief, that the high symmetry
may significantly simplify the structure of these equations. Indeed, it was
found in the leading logarithmic approximation (LLA)~\cite{BFKL2}, that the
so-called quasi-partonic operators in ${\mathcal N}=1$ SYM are unified in
supermultiplets with anomalous dimensions obtained from universal
anomalous dimensions $\gamma_{uni}(j)$ by shifting its arguments by an
integer number. Further, the anomalous dimension matrices for twist-2
operators are fixed by the superconformal invariance~\cite{BFKL2}.
Calculations in the maximally extended ${\mathcal N}=4$ SYM, where the
coupling constant is not renormalized, give even more remarkable
results. Namely, it turns out, that here all twist-2 operators enter in
the same multiplet, their anomalous dimension matrix is fixed completely
by the super-conformal invariance  and its universal anomalous dimension
in LLA is proportional to $\Psi (j-1)-\Psi (1)$, which means, that the
evolution equations for the matrix elements of quasi-partonic operators in
the multicolour limit $N_{c}\rightarrow \infty $ are equivalent to
the Schr\"{o}dinger equation for
an integrable Heisenberg spin model~\cite{N=4,LN4}. In QCD the
integrability remains only in a small sector of the quasi-partonic
operators~\cite{BDMB}. In the case of ${\mathcal N}=4$ SYM
the equations for other sets of operators are also
integrable~\cite{BS,DNW,MZ}. Evolution equations for quasi-partonic
operators are written in an explicitly super-conformal form in Ref.~\cite{RK}.

Similar results related
to the integrability of the multi-colour QCD were obtained
earlier in the Regge limit~\cite{Integr}. Moreover, it was shown~\cite{KL01},
that in the ${\mathcal N}=4$ SYM there is a deep relation between BFKL and DGLAP
evolution equations. Namely, the $j$-plane singularities of AD of the Wilson
twist-2 operators in this case can be obtained from the eigenvalues of the
BFKL kernel by their analytic continuation. The NLO
calculations in ${\mathcal N}=4$ SYM demonstrated~\cite{KL}, that some of these
relations are valid also in higher orders of perturbation theory. In
particular, the BFKL equation has the property of the hermitian
separability, the linear combinations of the multiplicatively renormalized
operators do not depend on the coupling constant, the eigenvalues of the
anomalous dimension matrix are expressed in terms of the universal
function $\gamma _{uni}(j)$ which can be obtained also from the BFKL
equation~\cite{KL}. The results for  $\gamma _{uni}(j)$ were checked
by the direct calculations in Ref.~\cite{KoLiVe}

In the ${\mathcal N}=4$ SYM theory~\cite{BSSGSO} we have the  following field content: one gluon
$g$, four Majorana fermions $\lambda $ and three complex scalars $\phi $. All
particles belong to the adjoint representation of the gauge group $SU(N_{c})$.
This model possesses an internal $SU(4)$ symmetry. In the ${\mathcal N}=4$ SYM theory
one can introduce the following colour and $SU(4)$ singlet local Wilson twist-2
operators~\cite{CPO,BFKL2,SUSYCWI,OV}:
\begin{eqnarray}
\mathcal{O}_{\mu _{1},...,\mu _{j}}^{g} &=&\hat{S}
G_{\rho \mu_{1}}^{a}{\mathcal D}_{\mu _{2}}
{\mathcal D}_{\mu _{3}}...{\mathcal D}_{\mu _{j-1}}G_{\rho \mu _{j}}^a\,,
\label{ggs}\\
{\tilde{\mathcal{O}}}_{\mu _{1},...,\mu _{j}}^{g} &=&\hat{S}
G_{\rho \mu_{1}}^a {\mathcal D}_{\mu _{2}}
{\mathcal D}_{\mu _{3}}...{\mathcal D}_{\mu _{j-1}}{\tilde{G}}_{\rho \mu _{j}}^a\,,
\label{ggp}\\
\mathcal{O}_{\mu _{1},...,\mu _{j}}^{\lambda } &=&\hat{S}
\bar{\lambda}_{i}^{a}\gamma _{\mu _{1}}
{\mathcal D}_{\mu _{2}}...{\mathcal D}_{\mu _{j}}\lambda ^{a\;i}\,, \label{qqs}\\
{\tilde{\mathcal{O}}}_{\mu _{1},...,\mu _{j}}^{\lambda } &=&\hat{S}
\bar{\lambda}_{i}^{a}\gamma _{5}\gamma _{\mu _{1}}{\mathcal D}_{\mu _{2}}...
{\mathcal D}_{\mu_{j}}\lambda ^{a\;i}\,, \label{qqp}\\
\mathcal{O}_{\mu _{1},...,\mu _{j}}^{\phi } &=&\hat{S}
\bar{\phi}_{r}^{a}{\mathcal D}_{\mu _{1}}
{\mathcal D}_{\mu _{2}}...{\mathcal D}_{\mu _{j}}\phi _{r}^{a}\,,\label{phphs}
\end{eqnarray}
where ${\mathcal D}_{\mu }$ are covariant derivatives. The spinors $\lambda _{i}$ and
field tensor $G_{\rho \mu }$ describe gluinos and gluons, respectively, and
$\phi _{r}$ are the complex scalar fields appearing in the ${\mathcal N}=4$ supersymmetric
model. Indices $i=1,\cdots ,4$ and $r=1,\cdots ,6$ refer to $SU(4)$ and
$SO(6)\simeq SU(4)$ groups of inner symmetry, respectively. The symbol
$\hat{S}$ implies a symmetrization of each tensor in the Lorentz indices $\mu
_{1},...,\mu _{j}$ and a subtraction of its traces. The anomalous dimension
matrices can be written for unpolarized and polarized cases, respectively,
as follows
\begin{equation}
\gamma _{\mathbf{unpol}}=
\begin{array}{|ccc|}
\gamma _{gg} & \gamma _{g\lambda } & \gamma _{g\phi } \\
\gamma _{\lambda g} & \gamma _{\lambda \lambda } & \gamma _{\lambda \phi } \\
\gamma _{\phi g} & \gamma _{\phi \lambda } & \gamma _{\phi \phi }
\end{array}
\ ,\qquad \qquad \gamma _{\mathbf{pol}}=
\begin{array}{|cc|}
{\tilde{\gamma}}_{gg} & {\tilde{\gamma}}_{g\lambda } \\
{\tilde{\gamma}}_{\lambda g} & {\tilde{\gamma}}_{\lambda \lambda }
\end{array}
\ .\label{MAD}
\end{equation}

Note, that in the super-multiplet of twist-2 operators there are also
operators with fermion quantum numbers and operators anti-symmetric in two
Lorentz indices~\cite{BFKL2,OV}.
For the case ${\mathcal N}=4$ the multiplicatively renormalized operators were
found in an explicit way and their universality properties for all orders of
perturbation theory were formulated in Refs.~\cite{LN4,KL}.

After their diagonalization, the new unpolarized $\gamma $ and polarized
$\tilde{\gamma}$ AD matrices have the following form
\begin{equation}
\gamma ~=~V^{-1}\gamma _{\mathbf{unpol}}V=
\begin{array}{|ccc|}
\gamma _{+}(j) & \gamma _{+0}(j) & \gamma _{+-}(j) \\
\gamma _{0+}(j) & \gamma _{0}(j) & \gamma _{0-}(j) \\
\gamma _{-+}(j) & \gamma _{-0}(j) & \gamma _{-}(j)
\end{array}
~,~~\tilde{\gamma}~=~\tilde{V}^{-1}\gamma _{\mathbf{pol}}\tilde{V}=
\begin{array}{|cc|}
{\tilde{\gamma}}_{+}(j) & {\tilde{\gamma}}_{+-}(j) \\
{\tilde{\gamma}}_{-+}(j) & {\tilde{\gamma}}_{-}(j)
\end{array}
\,,\label{maAD}
\end{equation}
which corresponds to AD matrices for multiplicatively renormalizable linear
combinations of operators~(\ref{ggs})-(\ref{phphs}). Here, the matrices $V,\ V^{-1},\ \tilde{V}$
and $\tilde{V}^{-1}$ were calculated in~\cite{KL} and in LO we have
$\gamma_{lm}(j)=0$, $\tilde{\gamma}_{lm}(j)=0$ for $l,m=+,0,-$.
In NLO the AD matrices become triangle~\cite{KoLiVe} due to superconformal
invariance breaking~\cite{OVJHEP}, similar to the case of ${\mathcal N}=1$
SYM~\cite{SUSYCWI}. The eigenvalues $\gamma _{l}(j)$ and $\tilde{\gamma}_{l}(j)$
govern the power-like violation of the Bjorken scaling for the parton
distributions.

Due to the fact that all twist-2 operators belong to the same supermultiplet
the anomalous dimensions $\gamma _{l}(j)$ and $\tilde{\gamma}_{l}(j)$
$(l=+,0,-)$ have the properties~\cite{LN4,KL}
\begin{equation}
\gamma _{+}(j)=\tilde{\gamma}_{+}(j-1)=\gamma _{0}(j-2)=
\tilde{\gamma}_{-}(j-3)=\gamma _{-}(j-4)=\gamma_{uni}(j),\label{Un}
\end{equation}
where $\gamma_{uni}(j)$ is the universal anomalous dimension.
\vskip 3mm
{\bf Method of obtaining AD eigenvalues in ${\mathcal N}=4$ SYM}
\vskip 3mm
As it was already pointed out in Introduction, the universal anomalous
dimension can be extracted directly from the QCD results without finding the
scalar particle contribution. This possibility is based on deep
relation between DGLAP and BFKL dynamics in the ${\mathcal N}=4$ SYM~\cite{KL01,KL}.

To begin with, the eigenvalues of the BFKL kernel turn out to be
analytic functions of the conformal spin $\left| n\right| $ at least in two first orders of
perturbation theory \cite{KL}. Further, in the framework of the ${\overline{\mathrm{DR}}}$-scheme~\cite{DRED}
one can obtain from the BFKL equation (see~\cite{KL01}), that there is no
mixing among the special functions of different transcendentality levels $i$
\footnote{%
Note that similar arguments were used also in~\cite{FleKoVe} to obtain
analytic results for contributions of some complicated massive Feynman
diagrams without direct calculations.},
i.e. all special functions at the NLO correction contain only sums of the
terms $\sim 1/j^{i}~(i=3)$. More precisely, if we introduce the
transcendentality level for the eigenvalues of integral kernels of the BFKL
equations as functions of $\gamma $ and appearing in the
perturbation theory in an accordance with the complexity of the terms in the
corresponding sums
\[
\Psi \sim 1/\gamma ,~~~\Psi ^{\prime }\sim \beta ^{\prime }\sim \zeta
(2)\sim 1/\gamma ^{2},~~~\Psi ^{\prime \prime }\sim \beta ^{\prime \prime
}\sim \zeta (3)\sim 1/\gamma ^{3},
\]
then for the BFKL kernel in the leading order (LO) and in NLO the
corresponding levels are $i=1$ and $i=3$, respectively.

Because in ${\mathcal N}=4$ SYM there is a relation between the BFKL and DGLAP equations
(see~\cite{KL01,KL}), the similar properties should be valid for the
anomalous dimensions themselves, i.e. the basic functions $\gamma
_{uni}^{(0)}(j)$, $\gamma _{uni}^{(1)}(j)$ and $\gamma _{uni}^{(2)}(j)$ are
assumed to be of the types $\sim 1/j^{i}$ with the levels $i=1$, $i=3$ and
$i=5$, respectively. An exception could be for the terms appearing at a given
order from previous orders of the perturbation theory. Such
contributions could be generated and/or removed by an approximate finite
renormalization of the coupling constant. But these terms do not appear in
the ${\overline{\mathrm{DR}}}$-scheme.

It is known, that at the LO and NLO approximations the
most complicated contributions (with $i=1$ and $i=3$, respectively) are the
same for all LO and NLO anomalous dimensions in QCD~\cite{LONLOAD}
and for the LO and NLO scalar-scalar anomalous
dimensions~\cite{KoLiVe}. This property allows one to find the
universal anomalous dimensions $\gamma _{uni}^{(0)}(j)$ and $\gamma
_{uni}^{(1)}(j)$ without knowing all elements of the anomalous dimension
matrix~\cite{KL}, which was verified by the exact calculations
in~\cite{KoLiVe}.

Using above arguments, we conclude, that at the NNLO level there is only one
possible candidate for $\gamma _{uni}^{(2)}(j)$. Namely, it is the most
complicated part of the nonsinglet QCD anomalous dimension matrix
(with the SUSY relation for the QCD color factors $C_{F}=C_{A}=N_{c}$).
Indeed, after the diagonalization of the
anomalous dimension matrix the eigenvalues $\gamma _{l}(j)$ and
$\tilde{\gamma}_{l}(j)$ in Eq.~(\ref{maAD}) should have this most complicated part
as a common contribution because they differ each from others only by a shift of
the argument (see Eq.~(\ref{Un})) and their differences are constructed from
less complicated terms. The non-diagonal matrix elements of $\gamma _{ab}(j)$
in Eq.~(\ref{maAD})
contain also only less complicated terms (see, for example, AD exact expressions
at LO and NLO approximations in Refs.~\cite{LONLOAD} for QCD
and~\cite{KoLiVe} for ${\mathcal N}=4$ SYM) and therefore they cannot generate
the most complicated contributions to $\gamma _{l}(j)$ and $\tilde{\gamma}_{l}(j)$.

Thus, the most complicated part of the nonsinglet NNLO QCD anomalous
dimension should coincide (up to color factors)
with the universal anomalous dimension $\gamma_{uni}^{(2)}(j)$.
\vskip 3mm
{\bf NNLO anomalous dimension for ${\mathcal N}=4$ SYM}
\vskip 3mm
The final three-loop result
\footnote{
Note, that in an accordance with Ref.~\cite{next}
 our normalization of $\gamma (j)$ contains
the extra factor $-1/2$ in comparison with
the standard normalization (see~\cite{LONLOAD})
and differs by sign in comparison with
Vermaseren-Moch-Vogt one~\cite{VeMoVo}.}
for the universal anomalous dimension $\gamma_{uni}(j)$
for ${\mathcal N}=4$ SYM is
\begin{eqnarray}
\gamma(j)\equiv\gamma_{uni}(j) ~=~ \hat a \gamma^{(0)}_{uni}(j)+\hat a^2
\gamma^{(1)}_{uni}(j) +\hat a^3 \gamma^{(2)}_{uni}(j) + ... , \qquad \hat a=%
\frac{\alpha N_c}{4\pi}\,,  \label{uni1}
\end{eqnarray}
where
\begin{eqnarray}
\frac{1}{4} \, \gamma^{(0)}_{uni}(j+2) &=& - S_1,  \label{uni1.1} \\
\frac{1}{8} \, \gamma^{(1)}_{uni}(j+2) &=& \Bigl(S_{3} +
\overline S_{-3} \Bigr) - 2\, \overline S_{-2,1} + 2\,S_1\Bigl(S_{2} +
 \overline S_{-2} \Bigr),  \label{uni1.2} \\
\frac{1}{32} \, \gamma^{(2)}_{uni}(j+2) &=& 2\, \overline S_{-3}\,S_2 -S_5 -
2\, \overline S_{-2}\,S_3 - 3\, \overline S_{-5}
+24\, \overline S_{-2,1,1,1}\nonumber\\
&&\hspace{-1.5cm}+ 6\biggl( \overline S_{-4,1} +  \overline S_{-3,2} +
 \overline S_{-2,3}\biggr)
- 12\biggl( \overline S_{-3,1,1} +  \overline S_{-2,1,2} +
\overline S_{-2,2,1}\biggr)\nonumber \\
&& \hspace{-1.5cm}  -
\biggl(S_2 + 2\,S_1^2\biggr) \biggl( 3 \, \overline S_{-3} + S_3
- 2\,  \overline S_{-2,1}\biggr)
- S_1\biggl(8\, \overline S_{-4} +  \overline S_{-2}^2\nonumber \\
&& \hspace{-1.5cm}  + 4\,S_2\, \overline S_{-2} +
2\,S_2^2 + 3\,S_4 - 12\,  \overline S_{-3,1} - 10\,  \overline S_{-2,2}
+ 16\,  \overline S_{-2,1,1}\biggr)
\label{uni1.5}
\end{eqnarray}
and $S_{a} \equiv S_{a}(j),\ S_{a,b} \equiv S_{a,b}(j),\ S_{a,b,c} \equiv
S_{a,b,c}(j)$ are harmonic sums
\begin{eqnarray}
&&\hspace*{-1cm} S_{a}(j)\ =\ \sum^j_{m=1} \frac{1}{m^a},
\ \ S_{a,b,c,\cdots}(j)~=~ \sum^j_{m=1}
\frac{1}{m^a}\, S_{b,c,\cdots}(m),  \label{ha1} \\
&&\hspace*{-1cm} S_{-a}(j)~=~ \sum^j_{m=1} \frac{(-1)^m}{m^a},~~
S_{-a,b,c,\cdots}(j)~=~ \sum^j_{m=1} \frac{(-1)^m}{m^a}\,
S_{b,c,\cdots}(m),  \nonumber \\
&&\hspace*{-1cm} \overline S_{-a,b,c,\cdots}(j) ~=~ (-1)^j \, S_{-a,b,c,...}(j)
+ S_{-a,b,c,\cdots}(\infty) \, \Bigl( 1-(-1)^j \Bigr).  \label{ha3}
\end{eqnarray}

The expression~(\ref{ha3}) is defined for all integer values of arguments (see~\cite{KK,KL})
but can be easily analytically continued to real and complex $j$
by the method of Refs.~\cite{AnalCont,KL}.

\vskip 3mm
{\bf The limit $j\rightarrow 1$ }
\vskip 3mm
The limit $j\rightarrow 1$ is important for the investigation of the small-$x
$ behavior of parton distributions (see review~\cite{Lund} and references
therein). Especially it became popular recently because there are
new experimental data at small $x$ produced by the H1 and ZEUS
collaborations in HERA~\cite{H1}.

Using asymptotic expressions for harmonic sums at $j=1+\omega \rightarrow 1$
(see formulae in Appendix at $r=-1$) we obtain for the ${\mathcal N}=4$ universal anomalous
dimension $\gamma_{uni}(j)$ in Eq.~(\ref{uni1})
\begin{eqnarray}
\gamma _{uni}^{(0)}(1+\omega ) &=&\frac{4}{\omega }+{\mathcal O}\Bigl(\omega^{1}\Bigr),
\label{uni4.1} \\
\gamma _{uni}^{(1)}(1+\omega ) &=&-32\,\zeta _{3}+{\mathcal O}\Bigl(\omega^{1}\Bigr),
\label{uni4.2} \\
\gamma _{uni}^{(2)}(1+\omega ) &=&32\zeta _{3}\,\frac{1}{\omega ^{2}}
-232\,\zeta _{4}\,\frac{1}{\omega }
-1120\zeta _{5}+256\zeta _{3}\zeta _{2}
+{\mathcal O}\Bigl(\omega ^{1}\Bigr)  \label{uni4.3}
\end{eqnarray}
in an agreement with the
predictions for $\gamma _{uni}^{(0)}(1+\omega )$,
$\gamma _{uni}^{(1)}(1+\omega )$ and also for the first term of
$\gamma _{uni}^{(2)}(1+\omega )$ coming from an investigation of BFKL
equation at NLO accuracy in~\cite{KL01}
\footnote{Unfortunately,
the results of Refs.~\cite{KL01,KL} contain a misprint.
Namely, the coefficient in front of $\hat{a}^3$ obtained in the limit
$j\rightarrow 1$ in Eq.~(39) of Ref.~\cite{KL} should be multiplied by a factor 4.}.

\vskip 3mm
{\bf The limit $j\to -r,\ r\geq 0 $}
\vskip 3mm

In the case of ${\mathcal N}=4$ SYM the
eigenvalue of the BFKL kernel~\cite{KL01}
is analytic in the conformal spin $|n|$, which allows
us to continue it to the
negative values of $|n|$~\cite{KL01}. It gives a possibility to find the
singular contributions to anomalous dimensions of the twist-2 operators not
only at $j=1$ but also at other integer negative points $j=0,\,-1,\,-2,
\cdots$.
In the Born approximation for the
universal anomalous dimension of the supermultiplet
of the twist-2 operators one can obtain in such way
$\gamma _{uni}=4\,\hat{a} \,
(\Psi (1)-\Psi (j-1))$~\cite{KL01}, which coincides with the result of the
direct calculations (see~\cite{N=4,LN4} and Eq.~(\ref{uni1.1})).

In the NLO approximation the study of the relation between BFKL and DGLAP
equation at $j\to -r,\ r\geq 0  $
was done in~\cite{KL}. Its extension to the NNLO level is a subject of
future investigations. Here we present only the AD results in this limit.

Using formulae presented in Appendix for the harmonic sums calculated in the
limit $j=\omega -r \rightarrow -r,\ r\geq 0$ we can find singularities of
$\gamma _{uni} (j)$~(\ref{uni1}) at small $\omega $
\begin{eqnarray}
\gamma _{uni}^{(0)}(j) &=&4\biggl[\frac{1}{\omega }-S_{1}(r+1)-
\Bigl(S_{2}(r+1)+\zeta_2 \Bigr) \omega +O(\omega ^{2})\biggr],
\label{dg2.4} \\
\gamma _{uni}^{(1)}(j) &=&8\biggl[\frac{(1+(-1)^{r})}{\omega ^{3}}
-2S_{1}(r+1)\frac{(1+(-1)^{r})}{\omega ^{2}}  \nonumber \\
&-&\Bigl((1+(-1)^{r})\zeta (2)+2(-1)^{r}S_{2}(r+1)\Bigr)\frac{1}{\omega }
+O(\omega ^{0})\biggr],
\label{dg2.5} \\
\gamma _{uni}^{(2)}(j) &=& 4\biggl[
\frac{c^{(5)}}{\omega^5}+\frac{c^{(4)}}{\omega^4}+\frac{c^{(3)}}{\omega^3}
+\frac{c^{(2)}}{\omega^2} +\frac{c^{(1)}}{\omega} + O(\omega ^0)\biggr],
\label{dg2.6}
\end{eqnarray}
where
\begin{eqnarray}
c^{(5)} &=& 16(1+(-1)^r), \\
c^{(4)} &=&  -32 (1+2(-1)^r) S_{1}, \\
c^{(3)} &=&  -8 \Biggl[  S_{2}-2 S_{1}^2-2 S_{-2}+4\zeta_2\Biggr]
 - 8(-1)^r \Biggl[5 S_{2}-6 S_{1}^2+4\zeta_2\Biggr], \\
c^{(2)} &=& -4\Biggl[ 4 S_{-2} S_{1}-8 S_{-3}
-2 S_{3}-6\zeta_2 S_{1}-7\zeta_3\Biggr] \nonumber \\
&& - 4(-1)^r\Biggl[4 S_{-2} S_{1}+2 S_{-3}
-16 S_{1} S_{2}-4 S_{-2,1}-18\zeta_2 S_1 -5\zeta_3 \Biggr], \\
c^{(1)} &=& \Biggl[
48 S_{-4} - 32 S_{-3} S_{1} - 16 S_{-2} S_{2}
+8 S_{-2}^2 \Biggr. \nonumber \\
&& \Biggl. -8 S_{4} - 24\zeta_2 S_{-2}
+8\zeta_2 S_{2} -24\zeta_3 S_{1} -3\zeta_4 \Biggr] \nonumber \\
&& +(-1)^r\Biggl[
8 S_{-4} + 32 S_{-3}S_{1}
+ 32 S_{1} S_{3}  \Biggr. \nonumber \\
&& + 56 S_{4} -32 S_{-3,1} -64 S_{1} S_{-2,1} + 32 S_{-2,2}
+ 32 S_{-2} S_{1}^2  \nonumber \\
&& \Biggl. + 16 S_{2}^2 + 64 S_{-2,1,1} + 8\zeta_2 S_{-2} + 40\zeta_2 S_{2}
- 32\zeta_2 S_{1}^2 -48\zeta_3 S_{1} + 55\zeta_4 \Biggr].
\end{eqnarray}
Here, to shorten notation when presenting resulting expression for singularities
of $\gamma _{uni}^{(2)}(j)$ we dropped argument $r+1$ in all harmonic sums.
So, the double-logarithmic poles
$\sim \omega ^{-3}$ and $\sim \omega ^{-5}$ appear in the anomalous
dimensions $\gamma _{uni}^{(1)}(j)$ and $\gamma _{uni}^{(2)}(j)$,
respectively, only in the case of even $r$ values, that is
in an agreement with
the predictions from the BFKL equation~\cite{KL}.

\vskip 3mm
{\bf Resummation of $\gamma_{uni}$ and the AdS/CFT correspondence}
\vskip 3mm
In the limit $j\to \infty $ the AD results~(\ref{uni1.1})-(\ref{uni1.5}) are
simplified significantly. Note, that this limit is related to the study of
the asymptotics of structure functions and cross-sections at $x\rightarrow 1$
corresponding to the quasi-elastic kinematics of the deep-inelastic $ep$
scattering.

We obtain the following
asymptotics for the ${\mathcal N}=4$ universal anomalous dimension $\gamma_{uni}(j)$ in
Eq.~(\ref{uni1}) with
\begin{eqnarray}
\gamma _{uni}^{(0)}(j) &=&-4\Bigl(\ln j+\gamma _{e}\Bigr)+{\mathcal O}\Bigl(j^{-1}
\Bigr),  \label{uni3.1} \\
\gamma _{uni}^{(1)}(j) &=&8\zeta _{2}\,\Bigl(\ln j+\gamma _{e}\Bigr)+12\zeta
_{3}+{\mathcal O}\Bigl(j^{-1}\Bigr),  \label{uni3.2} \\
\gamma _{uni}^{(2)}(j) &=&-88\zeta _{4}\,\Bigl(\ln j+\gamma _{e}
\Bigr)-16\zeta _{2}\zeta _{3}-80\zeta _{5}+{\mathcal O}\Bigl(j^{-1}\Bigr).  \label{uni3.3}
\end{eqnarray}

Recently there was a great progress in the investigation of the ${\mathcal N}=4$ SYM
theory in a framework of the AdS/CFT correspondence~\cite{2M} where the
strong-coupling limit $\alpha _{s}N_{c}\rightarrow \infty $ is described by
a classical supergravity in the anti-de Sitter space $AdS_{5}\times S^{5}$.
In particular, a very interesting prediction~\cite{15M} (see also~\cite{M})
was obtained for the large-$j$ behavior of the anomalous dimension for
twist-2 operators
\begin{equation}
\gamma (j)=a(z)\,\ln j \,,\qquad\qquad z=\frac{\alpha N_{c}}{\pi } =4\hat a
\end{equation}
in the strong coupling regime (see Ref.~\cite{16M} for asymptotic
corrections)\footnote{
Here we took into account, that in our normalization $\gamma (j)$
contains
the extra factor $-1/2$ in comparison with that in Ref.~\cite{15M}.}
:
\begin{equation}
\lim_{z\rightarrow \infty }a=-z^{1/2}+\frac{3\ln 2}{8 \pi}+{\mathcal O}
\left(z^{-1/2}\right) \,.
\label{1d}
\end{equation}

On the other hand, all anomalous dimensions $\gamma _{i}(j)$ and
$\tilde{\gamma}_{i}(j)$ ($i=+,0,-$) coincide at large $j$ and our results for
$\gamma_{uni}(j)$ in Eq.~(\ref{uni1}) allow one to find three first terms of the small-$z$
expansion of the coefficient $a(z)$
\begin{equation}
\lim_{z\rightarrow 0}\,a=-z+\frac{\pi ^2}{12}\, z^2-\frac{11}{720}
\pi^4z^3+...\,.
\end{equation}

For resummation of this series
we suggest the following equation for
$\tilde{a}$~\cite{KoLiVe}\footnote{Note, that we use the ${\overline{\mathrm{DR}}}$-scheme for
coupling constant which removes $-1/12$ from coefficients of $\widetilde a$ in Eq.(28) of Ref.~\cite{KoLiVe}
(see~\cite{KL,Belitsky:2003ys}).}
\begin{equation}
z=-\widetilde{a}+\frac{\pi ^{2}}{12}\,\widetilde{a}^{2}\,
\end{equation}
interpolating between its weak-coupling expansion up to NNLO
\begin{equation}
\tilde{a}=-z+\frac{\pi ^{2}}{12}\,z^{2}-\frac{1}{72}\pi ^{4}z^{3}+{\mathcal O}(z^{4})
\end{equation}
and strong-coupling asymptotics
\begin{equation}
\tilde{a} = -\frac{2\sqrt{3}}{\pi} \,z^{1/2} + \frac{6}{\pi^2} + {\mathcal O}
\left(z^{-1/2}\right)
\approx -1.1026\,\,z^{1/2}+0.6079+{\mathcal O}
\left(z^{-1/2}\right).
\end{equation}
It is remarkable, that the prediction for NNLO based on the above simple
equation is valid with the accuracy $\sim 10\%$. It means, that this
extrapolation seems to be good for all values of $z$.

Further, for $j\rightarrow 2$ let us take into account, that
according to the BFKL equation~\cite{next}
the anomalous dimension of twist-2 operators is quantized
in the Regge kinematics:
\begin{equation}
\gamma =1/2+i\nu +(j-1)/2= 1+(j-2)/2 +i\nu
\label{d1}
\end{equation}
 for the
principal series of unitary representations of the M\"{o}bius group.
%
On the other hand, in the diffusion
approximation valid near the leading singularity of the $t$-channel
partial wave the eigenvalue of the BFKL kernel is
\begin{equation}
j-1=\omega _0-D\nu^2\,,
\label{d2}
\end{equation}
where $\omega _0$ and $D$ are the Pomeron intercept and diffusion
coefficient, respectively. These quantities are functions of the
coupling constant. We assume, that for the large coupling constant
in $N=4$ SUSY the Pomeron intercept approaches the graviton
intercept in the $AdS_5\times S_5$ space ~\cite{polch}, which
means, that
\begin{equation}
j_0=1+\omega _0=2-\Delta \,,
\label{d3}
\end{equation}
where $\Delta$ is a small number.
Further, due to the energy-momentum conservation ($\gamma =0$ for
$j=2$) the parameters $\Delta$ and $D$ are equal and $\gamma (j)$
can be expressed near $j=2$ only in terms of one parameter
\begin{equation}
\gamma (j)=(j-2)\,
\left[ \frac{1}{2} - \frac{1/\Delta
}{1+
\sqrt{1+(j-2)/\Delta
}} \right].
\label{d4}
\end{equation}

The derivative $\gamma
^{\prime }(2)$ can be calculated from our  results in three first orders
of the
perturbation theory:
\begin{equation}
\gamma ^{\prime}(2) =
\frac{1}{2} - \frac{1}{2\Delta}
=-\frac{\pi ^2}{6}\,z+ \frac{\pi ^4}{72}
\,z^2- \frac{\pi ^6}{540}
\,z^3+...\,.
\label{gamma'}
\end{equation}
Similar to the case of large $j$ for a resummation of this series we used the
following
equation for $\widetilde{\widetilde a}=\gamma^{\prime }(2)$ (see \cite{KoLiVe})
\begin{equation}
\frac{\pi ^2}{6}\,
z=-
\, \widetilde{\widetilde a} +
\frac{1}{2} \,\widetilde{\widetilde a}^{2}\,.
\label{simple}
\end{equation}
Its solution at small $z$ is
\begin{equation}
\widetilde{\widetilde a}=-\frac{\pi ^2}{6} \, z + \frac{\pi ^4}{72}
\,z^2-  \frac{\pi ^6}{432}
\,z^3+...\,.
\label{atilde}
\end{equation}
One can
verify from eqs. (\ref{gamma'}) and (\ref{atilde}), that the prediction
for NNLO based on the simple
equation (\ref{simple})
is valid with the accuracy $\sim 20\%$. Therefore we can hope, that
this method of resummation gives us a good estimate also for the
behavior of $a$ at
large $z$
\begin{equation}
\gamma ^{\prime}(2)= 1- \sqrt{\frac{\pi^2}{3} \, z +1} \approx -
\frac{\pi}{\sqrt{3}} \,z^{1/2} +1 +  {\mathcal O}
\left(z^{-1/2}\right) \, .
\end{equation}

Thus, one obtains for the intercept of the Pomeron in $N=4$ SUSY
from the resummation (\ref{simple}) at
large $z$ the result
\begin{equation}
j= 2- \frac{\sqrt{3}}{2\pi} \,z^{-1/2} - \frac{3}{4\pi^2} \,z^{-1}
-{\mathcal O}\left(z^{-2}\right) .
\label{d5}
\end{equation}

On the other hand,
from eqs. (\ref{d1}) and  (\ref{d2}), using also the following
relation valid in ADS/CFT correspondence for the string energy at
$j$ close to 2 \cite{2M,15M}
\begin{equation}
E^2=(j+\Gamma)^2 - 4,~~\Gamma=-2\gamma,
\end{equation}
we obtain, that the BFKL equation in the diffusion approximation
(\ref{d1}) is equivalent to the equation for the leading Regge trajectory
in the superstring theory
\begin{equation}
j=2+\frac{\alpha'}{2}t,~~ t=E^2/R^2,~~ \alpha' =
\frac{R^2}{2}\Delta ,
\end{equation}
where $R$ is the radius of the anti-de-Sitter space.

It is naturally to expect that this Regge trajectory remains
approximately linear (up to corrections to diffusion approximation
of the BFKL equation) for all values of $t$ and $j$. We can
attempt to use expression (\ref{d4}) also for large $z$
\begin{equation}
\gamma(j)|_{z \to \infty} = -\sqrt{j-2} \Delta^{-1/2} . \label{d6}
\end{equation}

This relation is in an agreement with the prediction of A. Polyakov and
other authors \cite{2M,15M}\footnote{We
thank  Emery Sokatchev who drew our attention
to this result.}.
\begin{equation}
\gamma(j)|_{(z,j) \to \infty} = -\frac{1}{2} E = -\sqrt{\pi j}
z^{1/4} - \frac{3\sqrt{\pi }}{4} \frac{j^{3/2}}{z^{1/4}} + ... ,
\end{equation}
providing that
\begin{equation}
\Delta =  \frac{1}{\pi }  z^{-1/2}. \label{d7}
\end{equation}

This number coincides up to 15\% with the estimate $\Delta =
[\sqrt{3}/(2\pi)] z^{-1/2}$ obtained in (\ref{d5}) from the
resummation procedure (\ref{simple}). We can expect, that
expression (\ref{d6}) with parameter $\Delta$ calculated in
(\ref{d7}) gives the anomalous dimension of twist-2 operators for
$z \to \infty$ and  all $j$ (neglecting the nonlinearity effects).

Recently we were informed by Prof. Chung-I Tan, that the above
correction $\Delta$ to the graviton spin $j=2$ coincides in form
with that obtained in his unpublished work with R. Brower, J.
Polchinsky and M. Strassler from the AdS/CFT correspondence. We
thank him for helpful discussions.

\vskip 3mm
{\bf Conclusion}
\vskip 3mm
Thus, in this paper we constructed the anomalous dimension $\gamma _{uni}(j)$ for
the ${\mathcal N}=4$ supersymmetric gauge theory in the next-to-next-to-leading
approximation and verified its self-consistency in the Regge ($j\rightarrow 1
$) and quasi-elastic ($j\rightarrow \infty $) regimes. Our result
for universal anomalous dimension at $j=4$ could be used to determine the
anomalous dimension
of Konishi operator~\cite{Konishi:1983hf}
up to 3-loops. It is remarkable, that our results coincide
\footnote{We are grateful to Niklas Beisert and Matthias Staudacher for pointing
this to us} with corresponding predictions from dilatation operator approach
and integrability \cite{Beisert:2003tq,Beisert:2003ys}. The method,
developed for this construction, can be applied also to less symmetric cases of
${\mathcal N}=1,\ 2$ SYM and QCD, which are very important for phenomenological
applications. For the verification of the AdS/CFT correspondence the
calculations of the various physical quantities in ${\mathcal N}=4$ SYM attract a great
interest due to a possibility to develop non-perturbative approaches to QCD.

We demonstrated above that the expressions interpolating between the week and strong
regime work remarkably well both in limit $j\to\infty$ and $j\to 2$. The integrability of the evolution
equations for the quasi-partonic
operators in LLA~\cite{N=4,LN4} is an interesting property of ${\mathcal N}=4$ SYM which
should be verified on NLO and NNLO level. We hope to discuss these problems in
our future publications.

\vspace{10mm} {\large \textbf{Acknowledgments.}}\\[5mm]
This work was supported by the Alexander von Humboldt fellowship (A.V. K.),
Alexander von Humboldt prize (L.N. L.), the RFBR grant 04-02-17094,
RSGSS-1124.2003.2, INTAS 00-00-366, PHY-0244853 and DE-FG02-96ER41005.

\section*{Appendix A}

\label{App:A}

\setcounter{equation}0

For the special functions, contributing to our anomalous dimensions we have
the following formulae
\begin{eqnarray}
S_{k}(j-2) &=& -\frac{1}{\omega^k}-(-1)^{k} S_{k}- \sum^{\infty}_{l=1} \frac{
\Gamma(l+k)}{l! \Gamma(k)} \Bigl[(-1)^{k} S_{k+l}-(-1)^{l}\zeta_{k+l}
\Bigr], \\
\overline S_{-k}(j-2) &=& (-1)^{r+1}\biggl\{
\frac{1}{\omega^k} + \zeta_{-k} (1-(-1)^r) + (-1)^k S_{-k}  \nonumber \\
&&~~+ \sum^{\infty}_{l=1} \frac{\Gamma(l+k)}{l! \Gamma(k)} \Bigl[(-1)^{k}
S_{-(k+l)}+(-1)^{l}\zeta_{-(k+l)} \Bigr] \biggr\}, \\
\overline S_{-2,1,1,1}(j-2) &=& (-1)^{r+1}\frac{1}{\omega}
\biggl\{ \zeta_{4} + S_{-3,1} + S_{-2,2} - S_{-4} - S_{-2,1,1} \biggr\}, \\
\overline S_{-3,1,1}(j-2) &=& (-1)^{r+1}\frac{1}{\omega}
\biggl\{ \frac{\zeta_{3}}{\omega} - \frac{1}{4}\zeta_4
 + S_{-3,1} - S_{-4}(r+1) \biggr\}, \\
\overline S_{-2,1,1}(j-2) &=& (-1)^{r+1}\biggl\{ \frac{\zeta_3}{\omega} -
\frac{S_{-2,1}}{\omega} + \frac{S_{-3}}{\omega} - \frac{\zeta_4}{4}
+ S_{-2,1,1} -\zeta_2 S_{-2} - 2S_{-2,2} \biggr.
\nonumber \\
&&  \biggl.+ 4 S_{-4}
- 3S_{-3,1} + (1-(-1)^r)S_{-2,1,1}(\infty) \biggr\}, \\
\overline S_{-2,1}(j-2) &=& (-1)^{r+1}\biggl\{ \frac{1}{\omega}(\zeta_2-S_{-2}) -
\zeta_3 - \frac{5}{8}\zeta_3(1-(-1)^r)  + S_{-2,1} - 3S_{-3}
\biggr.  \nonumber \\
&& \biggl.+ \omega \biggl[ \zeta_2S_{-2} +
S_{-2,2}
+ 2S_{-3,1} - 6S_{-4} + \frac{33}{16}\zeta_4 \biggr] %
 \biggr\}, \\
\overline S_{-2,1,2}(j-2) &=& (-1)^{r+1} \frac{1}{\omega}
\biggl\{ \frac{1}{\omega}
(S_{-3}-S_{-2,1})
+ \frac{5}{4}\zeta_4 -2S_{-3,1} -
S_{-2,2} + 3S_{-4} \biggr\}, \\
\overline S_{-2,2,1}(j-2) &=& (-1)^{r+1}\frac{1}{\omega}\biggl\{
\frac{7}{4}\zeta_4 +
S_{-2,2} - \zeta_2 S_{-2} - S_{-4} \biggr\}, \\
\overline S_{-3,2}(j-2) &=& (-1)^{r+1} \frac{1}{\omega}
\biggl\{ \frac{1}{\omega}\Bigl( S_{-3} + 2\zeta_3\Bigr) +
3S_{-4} - 3\zeta_{4} \biggr\}, \\
\overline S_{-2,2}(j-2) &=& (-1)^{r+1}\biggl\{ -\frac{1}{\omega^2}S_{-2}
- \frac{2}{\omega} S_{-3} -2S_{-4} -S_{-2,2}+ \frac{2}{\omega}\zeta_3 -3\zeta_4  \biggr.
\nonumber \\
&& \biggl. + (1-(-1)^r)S_{-2,2}(\infty)
\biggr\}, \\
\overline S_{-4,1}(j-2) &=& (-1)^{r+1} \frac{1}{\omega}
\biggl\{ -S_{-4} + \zeta_4 - \frac{\zeta_3}{\omega} + \frac{\zeta_2}{\omega^2}
\biggr\}, \\
\overline S_{-2,3}(j-2) &=& (-1)^{r+1} \frac{1}{\omega}
\biggl\{ -\frac{1}{\omega^2}S_{-2} - \frac{2}{\omega}S_{-3} - 3S_{-4} +
3\zeta_4 \biggr\}, \\
\overline S_{-3,1}(j-2) &=& (-1)^{r+1}\biggl\{ \frac{1}{\omega}S_{-3} +
\frac{\zeta_2}{\omega^2} - \frac{\zeta_3}{\omega} + \zeta_4 + 4S_{-4} - S_{-3,1}
\biggr.  \nonumber \\
&& \biggl.  + (1-(-1)^r)S_{-3,1}(\infty)
\biggr\},
\end{eqnarray}
where $\zeta_{-k}= (1/2^{k-1}-1)\zeta_{k}$ and the functions
$S_{a,b,c} \equiv S_{a,b,c}(r+1)$.

\end{document}